\documentclass[prb,twocolumn,floatfix,superscriptaddress]{revtex4}

\usepackage{graphicx}
\usepackage{amsmath}
\usepackage{amssymb}
\usepackage[colorlinks=true,citecolor=blue,linkcolor=blue]{hyperref}
\usepackage{amsfonts}
\usepackage{color,xcolor}
\usepackage{epstopdf}
\usepackage{braket}
\usepackage{bm}
\usepackage{bbold}

\renewcommand{\vec}[1]{\ensuremath{\boldsymbol{#1}}}

\begin{document}

\title{Rich many-body phase diagram of electrons and holes in doped monolayer transition metal dichalcogenides}
\date{\today}
\author{M. Van der Donck}
\email{matthias.vanderdonck@uantwerpen.be}
\affiliation{Department of Physics, University of Antwerp, Groenenborgerlaan 171, B-2020 Antwerp, Belgium}
\author{F. M. Peeters}
\email{francois.peeters@uantwerpen.be}
\affiliation{Department of Physics, University of Antwerp, Groenenborgerlaan 171, B-2020 Antwerp, Belgium}

\begin{abstract}
We use a variational technique to study the many-body phase diagram of electrons and holes in $n$-doped and $p$-doped monolayer transition metal dichalcogenides (TMDs). We find a total of four different phases. $i$) A fully spin polarized and valley polarized ferromagnetic state. $ii$) A state with no global spin polarization but with spin polarization in each valley separately, i.e. spin-valley locking. $iii$) A state with spin polarization in one of the valleys and little to no spin polarization in the other valley. $iv$) A paramagnetic state with no valley polarization. These phases are separated by first-order phase transitions and are determined by the particle density and the dielectric constant of the substrate. We find that in the presence of a perpendicular magnetic field the four different phases persist. In the case of $n$-doped MoS$_2$, a fifth phase, which is completely valley polarized but not spin polarized, appears for magnetic fields larger than 7 T and for magnetic fields larger than 23 T completely replaces the second phase.
\end{abstract}

\maketitle

\section{Introduction}

It is known that long-range exchange interactions cause the three-dimensional electron gas to become ferromagnetic at low densities\cite{ferro1,ferro2,ferro3}. This was later confirmed by Monte Carlo simulations\cite{ferro4,ferro5}, which predicted the same effect in the two-dimensional (2D) electron gas\cite{ferro4,ferro6}. The 2D electron gas has been investigated in detail in semiconductor heterostructures and electrons above liquid helium\cite{helium1,helium2,helium3}, however the discovery of graphene\cite{graphenedisc}, later followed by a whole range of different 2D materials, provided new systems with different dispersion relations and topologies for studying the 2D electron gas\cite{graphene2deg}. The pronounced effect of the dispersion relation on the many-body state of the electron gas was shown by using a variational wave function technique, which found that monolayer graphene does not exhibit a ferromagnetic phase\cite{netomono} while bilayer graphene does\cite{netobi}.

Another class of 2D materials is formed by the monolayer transition metal dichalcogenides (TMDs), such as MoS$_2$, MoSe$_2$, WS$_2$, WSe$_2$, etc.\cite{mak1,splendiani,mak2,zeng,cao,ross}. As opposed to graphene, monolayer TMDs lack inversion symmetry, which leads to a large direct band gap in the low-energy valleys at the corners of the first Brillouin zone. Furthermore, they exhibit a strong spin-orbit coupling, which leads to a large splitting of the valence bands and a small splitting of the conduction bands, which is opposite in the two valleys. It is expected that this valley-contrasting spin splitting will result in a rich many-body phase diagram with many more possible phases than those predicted for monolayer and bilayer graphene.

Recently, ferromagnetic behavior was predicted in numerous different TMD-based systems such as exfoliated TMDs with defects\cite{exfoliated}, transition metal-doped TMDs\cite{doped}, intercalated TMDs\cite{intercalated}, TMD-based heterostructures\cite{hetero1,hetero2}, and TMDs in which one of the chalcogen layers is either removed\cite{chalc1} or different from the other chalcogen layer\cite{chalc2}. However, in all of these systems the ferromagnetic phase is not driven by many-body exchange interactions but rather is a single-particle effect in which one of the spin states is energetically preferred over the other, and which is not present in clean monolayer TMDs.

In the present paper we use a variational technique, similar to that used in Refs. [\onlinecite{netomono,netobi}], to study the exchange interaction-driven many-body phase of different monolayer TMDs and its dependence on the dielectric constant of the substrate and on a perpendicular magnetic field. Our paper is organized as follows. In Sec. \ref{sec:Model} we present an outline of the theoretical model, including the many-body Hamiltonian and the variational state. The numerical results are discussed in Sec. \ref{sec:Results}. The main conclusions are summarized in Sec. \ref{sec:Summary and conclusion}.

\section{Model}
\label{sec:Model}

\subsection{Many-body Hamiltonian}

\subsubsection{Kinetic energy}

The effective low-energy single-particle kinetic Hamiltonian of monolayer TMDs is given by\cite{theory}
\begin{equation}
\label{singleham}
H_{\vec{k},\sigma,\tau} =
\begin{pmatrix}
\frac{\Delta}{2}+\lambda_c\sigma\tau & at(\tau k_x-ik_y) \\
at(\tau k_x+ik_y) & -\frac{\Delta}{2}+\lambda_v\sigma\tau
\end{pmatrix},
\end{equation}
with $a$ the lattice constant, $t$ the hopping parameter, $\sigma=\pm1$ the spin index, $\tau=\pm1$ the valley index, $\Delta$ the band gap, and $\lambda_c$ ($\lambda_v$) the spin-orbit coupling strength leading to a spin splitting of $2\lambda_c$ ($2\lambda_v$) at the conduction (valence) band edge. The values of these constants are listed in Table \ref{table:mattable} for different TMDs. The corresponding single-particle kinetic energy spectrum is given by
\begin{equation}
\label{singleenergy}
E_{\vec{k},\sigma,\tau,\pm} = \frac{\lambda_c+\lambda_v}{2}\sigma\tau\pm\sqrt{a^2t^2k^2+\frac{\Delta_{\sigma,\tau}^2}{4}},
\end{equation}
with $\Delta_{\sigma,\tau}=\Delta+(\lambda_c-\lambda_v)\sigma\tau$ and with the plus (minus) sign describing the conduction (valence) band. Because spontaneous electron-hole creation is suppressed due to the large band gap we have to consider either only conduction band states ($n$-doped TMDs) or only valence band states ($p$-doped TMDs). The many-body kinetic Hamiltonian can then be written as
\begin{equation}
\label{manyham}
\hat{H}_0 = \pm\sum_{\vec{k},\sigma,\tau}E_{\vec{k},\sigma,\tau,\pm}\hat{a}^{\dag}_{\vec{k},\sigma,\tau}\hat{a}_{\vec{k},\sigma,\tau},
\end{equation}
with the plus (minus) sign describing electrons (holes) and with $\hat{a}^{\dag}_{\vec{k},\sigma,\tau}$ ($\hat{a}_{\vec{k},\sigma,\tau}$) the creation (annihilation) operator of either an electron in the conduction band ($n$-doped TMDs) or a vacancy in the valence band ($p$-doped TMDs) with wave vector $\vec{k}$, spin index $\sigma$, and valley index $\tau$. The hole quasiparticle has opposite energy, momentum, spin index, and valley index as compared to that of the vacant valence band state.
\begin{table}
\centering
\caption{Lattice constant ($a$)\cite{theory}, hopping parameter ($t$)\cite{theory}, band gap ($\Delta$)\cite{theory}, spin splitting of the conduction ($2\lambda_c$)\cite{lambdac} and valence ($2\lambda_v$)\cite{lambdav} band, and 2D polarizability ($\chi_{\text{2D}}$) \cite{screeninglength} for different TMD materials.}
\begin{tabular}{c c c c c c c}
\hline
\hline
 & $a$ (nm) & $t$ (eV) & $\Delta$ (eV) & $2\lambda_c$ (eV)& $2\lambda_v$ (eV) & $\chi_{\text{2D}}$ (nm) \\
\hline
\hline
Mo$\text{S}_2$ & 0.32 & 1.10 & 1.66 & $-$0.003 & 0.15 & 8.29 \\
\hline
MoS$\text{e}_2$ & 0.33 & 0.94 & 1.47 & $-$0.021 & 0.18 & 10.34 \\
\hline
W$\text{S}_2$ & 0.32 & 1.37 & 1.79 & 0.027 & 0.43 & 7.58 \\
\hline
WS$\text{e}_2$ & 0.33 & 1.19 & 1.60 & 0.038 & 0.46 & 9.02 \\
\hline
\hline
\end{tabular}
\label{table:mattable}
\end{table}

\subsubsection{Interparticle interactions}

The interaction potential in monolayer TMDs is, due to non-local screening effects, given by\cite{screening1,screening2,screening3}
\begin{equation}
\label{interpot}
V(r) = \frac{e^2}{4\pi\kappa\varepsilon_0}\frac{\pi}{2r_0}\left[H_0\left(\frac{r}{r_0}\right)-Y_0\left(\frac{r}{r_0}\right)\right],
\end{equation}
with $r$ the distance between two particles, where $Y_0$ and $H_0$ are the Bessel function of the second kind and the Struve function, respectively, with $\kappa=(\varepsilon_b+\varepsilon_t)/2$ where $\varepsilon_{b(t)}$ is the dielectric constant of the environment below (above) the TMD monolayer, and with $r_0=\chi_{2\text{D}}/(2\kappa)$ the screening length where $\chi_{2\text{D}}$ is the 2D polarizability of the TMD. For $r_0=0$ this potential reduces to the bare Coulomb potential $V(r)=e^2/(4\pi\kappa\varepsilon_0r)$. Increasing the screening length leads to a decrease in the short-range interaction strength while the long-range interaction strength is unaffected. For very large screening lengths $r_0\rightarrow\infty$ the interaction potential becomes logarithmic, i.e. $V(r)=e^2/(4\pi\kappa\varepsilon_0r_0)\text{ln}(r_0/r)$. The $q$-dependence of the interaction potential \eqref{interpot} goes as $(q+r_0q^2)^{-1}$. The many-body interaction Hamiltonian is in general given by
\begin{widetext}
\begin{equation}
\label{manyinter}
\hat{V} = \frac{1}{2}\sum_{\vec{q},\nu,\rho}\sum_{\vec{k},\sigma,\tau}\sum_{\vec{q}',\nu',\rho'}\sum_{\vec{k}',\sigma',\tau'}\braket{\psi_{\vec{q},\nu,\rho}\psi_{\vec{k},\sigma,\tau}|V(r)|\psi_{\vec{q}',\nu',\rho'}\psi_{\vec{k}',\sigma',\tau'}}\hat{a}^{\dag}_{\vec{q},\nu,\rho}\hat{a}^{\dag}_{\vec{k},\sigma,\tau}\hat{a}_{\vec{k}',\sigma',\tau'}\hat{a}_{\vec{q}',\nu',\rho'},
\end{equation}
\end{widetext}
with $\ket{\psi_{\vec{k},\sigma,\tau}}$ the eigenstates of the Hamiltonian \eqref{singleham}
\begin{equation}
\label{eigentoestand}
\braket{\vec{r}|\psi_{\vec{k},\sigma,\tau}} = \sqrt{\frac{\beta_{\vec{k},\sigma,\tau}^2}{A\left(\beta_{\vec{k},\sigma,\tau}^2+a^2t^2k^2\right)}}
\begin{pmatrix}
1 \\ \frac{atk\tau e^{i\tau\theta}}{\beta_{\vec{k},\sigma,\tau}}
\end{pmatrix}
e^{i\vec{k}.\vec{r}}\eta_{\sigma},
\end{equation}
with
\begin{equation}
\label{normering}
\beta_{\vec{k},\sigma,\tau} = \frac{\Delta_{\sigma,\tau}}{2}\pm\sqrt{a^2t^2k^2+\frac{\Delta_{\sigma,\tau}^2}{4}},
\end{equation}
$\theta=\arctan(k_y/k_x)$, $A$ the surface area, $\eta_{\sigma}$ the orthonormal spin states, and with the plus (minus) sign describing electrons (holes). We then have
\begin{widetext}
\begin{equation}
\label{overlap}
\begin{split}
\braket{\psi_{\vec{q},\nu,\rho}\psi_{\vec{k},\sigma,\tau}|V(r)|\psi_{\vec{q}',\nu',\rho'}\psi_{\vec{k}',\sigma',\tau'}} = \frac{e^2}{2\varepsilon_0\kappa A}\delta_{\nu,\nu'}\delta_{\sigma,\sigma'}\delta_{\rho,\rho'}\delta_{\tau,\tau'}\delta_{\vec{k}'+\vec{q}',\vec{k}+\vec{q}}\frac{\braket{\psi_{\vec{q},\nu,\rho}|\psi_{\vec{q}',\nu',\rho'}}_p\braket{\psi_{\vec{k},\sigma,\tau}|\psi_{\vec{k}',\sigma',\tau'}}_p}{|\vec{k}-\vec{k}'|+\frac{\chi_{\text{2D}}}{2\kappa}|\vec{k}-\vec{k}'|^2},
\end{split}
\end{equation}
\end{widetext}
where $\braket{}_p$ denotes the overlap element of the pseudospin part of the eigenstates and where we have neglected intervalley scattering due to the large corresponding momentum exchange. When only considering the exchange interactions (the direct interactions should be canceled by the interactions with the positive lattice background), we find
\begin{equation}
\label{exchange}
\begin{split}
\hat{V} = -\frac{e^2}{4\varepsilon_0\kappa A}\sum_{\sigma,\tau}\sum_{\vec{k},\vec{q}}&\frac{|\braket{\psi_{\vec{q},\sigma,\tau}|\psi_{\vec{k},\sigma,\tau}}_p|^2}{|\vec{k}-\vec{q}|+\frac{\chi_{\text{2D}}}{2\kappa}|\vec{k}-\vec{q}|^2} \\
&\hat{a}^{\dag}_{\vec{q},\sigma,\tau}\hat{a}_{\vec{q},\sigma,\tau}\hat{a}^{\dag}_{\vec{k},\sigma,\tau}\hat{a}_{\vec{k},\sigma,\tau}.
\end{split}
\end{equation}

\subsubsection{Zeeman effect}

The presence of a perpendicular magnetic field leads to three different Zeeman effects by coupling with three different magnetic moments. $i$) The magnetic moments of the particles around their atomic site. The magnetic quantum numbers in the conduction and valence bands are given by $m_z=0$ and $m_z=2\tau$, respectively. $ii$) The spin magnetic moments of the particles. $iii$) The intrinsic magnetic moment of the individual Bloch particles\cite{blochmagn}
\begin{equation}
\label{magnmoment}
\begin{split}
\vec{m}_{\vec{k},\sigma,\tau} &= -i\frac{e}{2\hbar}\braket{\vec{\nabla}_{\vec{k}}\psi_{\vec{k},\sigma,\tau}|\times\left(H_{\vec{k},\sigma,\tau}-E_{\vec{k},\sigma,\tau}\right)|\vec{\nabla}_{\vec{k}}\psi_{\vec{k},\sigma,\tau}}_p \\
&= -\tau\frac{ea^2t^2\Delta_{\sigma,\tau}}{4\hbar a^2t^2k^2+\hbar\Delta_{\sigma,\tau}^2}\vec{e}_z.
\end{split}
\end{equation}
Putting this all together we get for the magnetic part of the many-body Hamiltonian
\begin{equation}
\label{manymagn}
\begin{split}
\hat{H}_B = &\pm B\sum_{\vec{k},\sigma,\tau}(\sigma\mu_B-\vec{e}_z.\vec{m}_{\vec{k},\sigma,\tau})\hat{a}^{\dag}_{\vec{k},\sigma,\tau}\hat{a}_{\vec{k},\sigma,\tau} \\
&-2B\sum_{\vec{k},\sigma,\tau}\tau\mu_B\hat{a}^{\dag}_{\vec{k},\sigma,\tau}\hat{a}_{\vec{k},\sigma,\tau}\ ,
\end{split}
\end{equation}
with the plus (minus) sign describing electrons (holes) and where the last term should only be included for holes.

Apart from the different Zeeman effects, a perpendicular magnetic field also leads to confinement of the charge carriers, which results in discrete Landau levels in the energy spectrum. This will have a significant effect on the many body phase when the confinement region is smaller than the average interparticle distance. The latter can be estimated by $\braket{r}=1/\sqrt{\pi n}$ with $n$ the charge carrier density, while the former is given by the magnetic length $l_B=\sqrt{\hbar/(eB)}$. For $B=50$ T we have $l_B=36.3$ \AA, which is less than the average interparticle distance for densities smaller than $n=0.3\times10^{13}$ cm$^{-2}$. This means that only at strong magnetic field strengths and low densities would the Landau levels significantly affect the many-body phase. Therefore, we do not take this effect into account in the current work.

\subsection{Variational solution}

We consider a variational state in which the four energy bands can be filled independently from each other up to a certain number of particles $N_{\sigma,\tau}$, i.e. the state
\begin{equation}
\label{grond}
\ket{\Psi_0} = \left(\prod_{\sigma,\tau}\prod_{k\leq k_F^{\sigma,\tau}}\hat{a}^{\dag}_{\vec{k},\sigma,\tau}\right)\ket{\emptyset},
\end{equation}
with $k_F^{\sigma,\tau}$ the band dependent Fermi wave vector and with $\ket{\emptyset}$ the vacuum state, i.e. completely filled valence bands and completely empty conduction bands. This is a Hartree-Fock method in which Fermi correlation is taken into account (through the anticommutation relations of the creation and annihilation operators) but Coulomb correlation isn't. The occupation number of a given single-particle state is therefore given by
\begin{equation}
\label{occupation}
N_{\vec{k},\sigma,\tau} = \braket{\Psi_0|\hat{a}^{\dag}_{\vec{k},\sigma,\tau}\hat{a}_{\vec{k},\sigma,\tau}|\Psi_0} =
\begin{cases}
1\text{ for }k\leq k_F^{\sigma,\tau} \\
0\text{ for }k>k_F^{\sigma,\tau}
\end{cases}.
\end{equation}
The total number of particles in a given energy band is given by $N_{\sigma,\tau}=\sum_{\vec{k}}N_{\vec{k},\sigma,\tau}$ and together they form the set of variational parameters. In order to gain more direct physical insight from the variational parameters we transform them to
\begin{equation}
\label{varpars}
\begin{split}
&N = \sum_{\sigma,\tau}N_{\sigma,\tau}, \hspace{36pt} \zeta_{\sigma} = \frac{\sum_{\sigma,\tau}\sigma N_{\sigma,\tau}}{\sum_{\sigma,\tau}N_{\sigma,\tau}}, \\
&\zeta_{\tau} = \frac{\sum_{\sigma,\tau}\tau N_{\sigma,\tau}}{\sum_{\sigma,\tau}N_{\sigma,\tau}}, \hspace{20pt} \zeta_{\alpha} = \frac{\sum_{\sigma,\tau}\sigma\tau N_{\sigma,\tau}}{\sum_{\sigma,\tau}N_{\sigma,\tau}}.
\end{split}
\end{equation}
The total number of particles in the system $N$ is fixed, meaning that we have three variational parameters: $\zeta_{\sigma}$, $\zeta_{\tau}$, and $\zeta_{\alpha}$. These are the spin, valley, and spin-valley polarization, respectively, and can range from $-1$ to $1$. For example, a state characterized by $(\zeta_{\sigma},\zeta_{\tau},\zeta_{\alpha})=(0,0,1)$ has an equal number of spin up and spin down particles, has an equal number of particles in both valleys, but all spin up (spin down) particles reside in the $K$ ($K'$) valley. Starting from $N_{\sigma,\tau}=\sum_{\vec{k}}N_{\vec{k},\sigma,\tau}$, converting the summation over $\vec{k}$ to an integral, and using Eq. \eqref{occupation} we obtain an expression relating $N_{\sigma,\tau}$ and $k_F^{\sigma,\tau}$. We then invert the set of equations in Eq. \eqref{varpars} to get
\begin{equation}
\label{fermiwave}
k_F^{\sigma,\tau} = \sqrt{4\pi\frac{N_{\sigma,\tau}}{A}} = \sqrt{\pi n(1+\sigma\zeta_{\sigma}+\tau\zeta_{\tau}+\sigma\tau\zeta_{\alpha})},
\end{equation}
with $n=N/A$ the total particle density.
\begin{figure*}
\centering
\includegraphics[width=17cm]{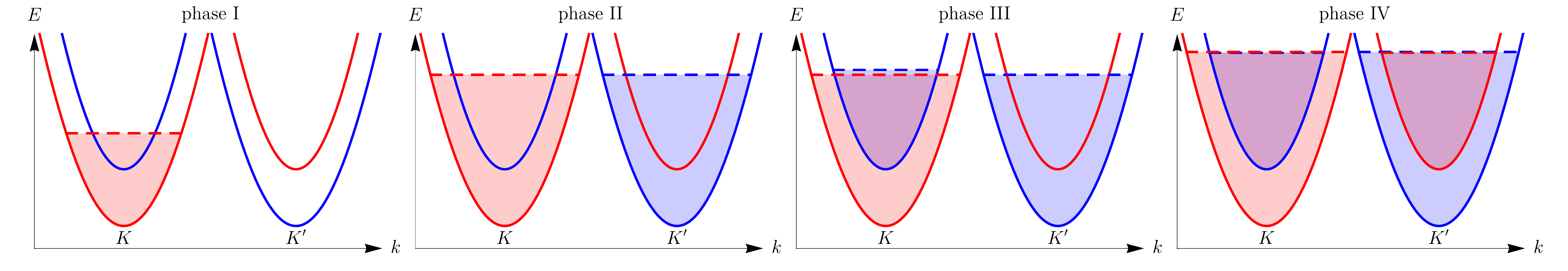}
\caption{(Color online) The four different phases of monolayer TMDs in zero magnetic field. Blue and red bands are spin up and spin down bands, respectively.}
\label{fig:phaseplot}
\end{figure*}

\subsubsection{Kinetic energy}

The expectation value of the kinetic energy \eqref{manyham} for our variational state is given by
\begin{equation}
\label{kineticexp}
\begin{split}
T &= \braket{\Psi_0|\hat{H}_0|\Psi_0} = \pm\frac{A}{2\pi}\sum_{\sigma,\tau}\int_0^{k_F^{\sigma,\tau}}dkkE_{\vec{k},\sigma,\tau,\pm} \\
&= \frac{N}{2\pi n}\sum_{\sigma,\tau}\bigg(\pm\frac{\lambda_c+\lambda_v}{2}\sigma\tau\left(k_F^{\sigma,\tau}\right)^2 \\
&\hspace{31pt}+\frac{1}{24a^2t^2}\left(\left(4a^2t^2\left(k_F^{\sigma,\tau}\right)^2+\Delta_{\sigma,\tau}^2\right)^{\frac{3}{2}}-\Delta_{\sigma,\tau}^3\right)\bigg).
\end{split}
\end{equation}

\subsubsection{Interparticle interactions}

The expectation value of the interparticle interactions \eqref{exchange} for our variational state is given by
\begin{equation}
\label{interexp}
V_C = \braket{\Psi_0|\hat{V}|\Psi_0} = -\frac{Ne^2}{4\varepsilon_0\kappa(2\pi)^3n}\sum_{\sigma,\tau}I_{\sigma,\tau}\left(k_F^{\sigma,\tau}\right)^3,
\end{equation}
where the integral 
\begin{widetext}
\begin{equation}
\label{integral}
I_{\sigma,\tau} = \int_0^1du\int_0^1dv\int_0^{2\pi}d\theta\frac{uv\left(u^2v^2+f_{\sigma,\tau}^2(u)f_{\sigma,\tau}^2(v)+2uvf_{\sigma,\tau}(u)f_{\sigma,\tau}(v)\cos\theta\right)}{4\left(u^2+\tilde{\Delta}_{\sigma,\tau}f_{\sigma,\tau}(u)\right)\left(v^2+\tilde{\Delta}_{\sigma,\tau}f_{\sigma,\tau}(v)\right)\left(\sqrt{u^2+v^2-2uv\cos\theta}+c_{\sigma,\tau}\left(u^2+v^2-2uv\cos\theta\right)\right)},
\end{equation}
\end{widetext}
with
\begin{equation}
\label{help}
\begin{split}
&f_{\sigma,\tau}(x) = \tilde{\Delta}_{\sigma,\tau}\pm\sqrt{x^2+\tilde{\Delta}_{\sigma,\tau}^2}, \\
&\tilde{\Delta}_{\sigma,\tau} = \frac{\Delta_{\sigma,\tau}}{2atk_F^{\sigma,\tau}}, \hspace{20pt} c_{\sigma,\tau} = \frac{\chi_{\text{2D}}k_F^{\sigma,\tau}}{2\kappa},
\end{split}
\end{equation}
is evaluated numerically.

\subsubsection{Zeeman effect}

The expectation value of the magnetic part of the Hamiltonian \eqref{manymagn} for our variational state is given by
\begin{equation}
\label{magnexp}
\braket{\hat{H}_B} = \pm N\zeta_{\sigma}\mu_BB\pm\frac{NeB}{16\pi\hbar n}\sum_{\sigma,\tau}\tau\Delta_{\sigma,\tau}\ln\left(1+\frac{1}{\tilde{\Delta}_{\sigma,\tau}^2}\right),
\end{equation}
where in the first term we have to substitute $\zeta_{\sigma}\rightarrow\zeta_{\sigma}+2\zeta_{\tau}$ for holes.

The sum of the three terms \eqref{kineticexp}, \eqref{interexp}, and \eqref{magnexp} gives the total variational energy, which depends on the three variational parameters. We minimize the variational energy brute force to find the variational parameters which define the lowest energy many-body state.

\section{Results}
\label{sec:Results}

The main results and discussions presented here are for $n$-doped TMDs, which are our main focus. In the next subsection we briefly comment on the results for $p$-doped TMDs.

\subsection{$n$-doped TMDs}

In the absence of interactions the many-body state can simply be found by filling up the lowest energy single-particle states. In the absence of a magnetic field this means that both valleys are populated equally and, as a consequence, that both spin states are also populated equally. The many-body state is therefore characterized by $(\zeta_{\sigma},\zeta_{\tau},\zeta_{\alpha})=(0,0,1)$ at low densities, i.e. there is no global spin and no valley polarization but there is spin polarization in each valley separately (spin-valley locking). For densities above some critical value, the electrons will also populate the higher conduction band in both valleys and as such the spin-valley locking will be gradually lost, i.e. there is a second-order phase transition. The many-body state is then given by $(\zeta_{\sigma},\zeta_{\tau},\zeta_{\alpha})=(0,0,\alpha(n))$ for Mo-based TMDs and by $(\zeta_{\sigma},\zeta_{\tau},\zeta_{\alpha})=(0,0,-\alpha(n))$ for W-based TMDs with
\begin{equation}
\label{zalphamin}
\alpha(n) = \frac{\Delta(\lambda_v-\lambda_c)-(\lambda_v+\lambda_c)\sqrt{4a^2t^2\pi n+\Delta^2-4\lambda_c\lambda_v}}{2a^2t^2\pi n}
\end{equation}
a function which decreases continuously with increasing density from 1 to 0.
\begin{figure}
\centering
\includegraphics[width=8.5cm]{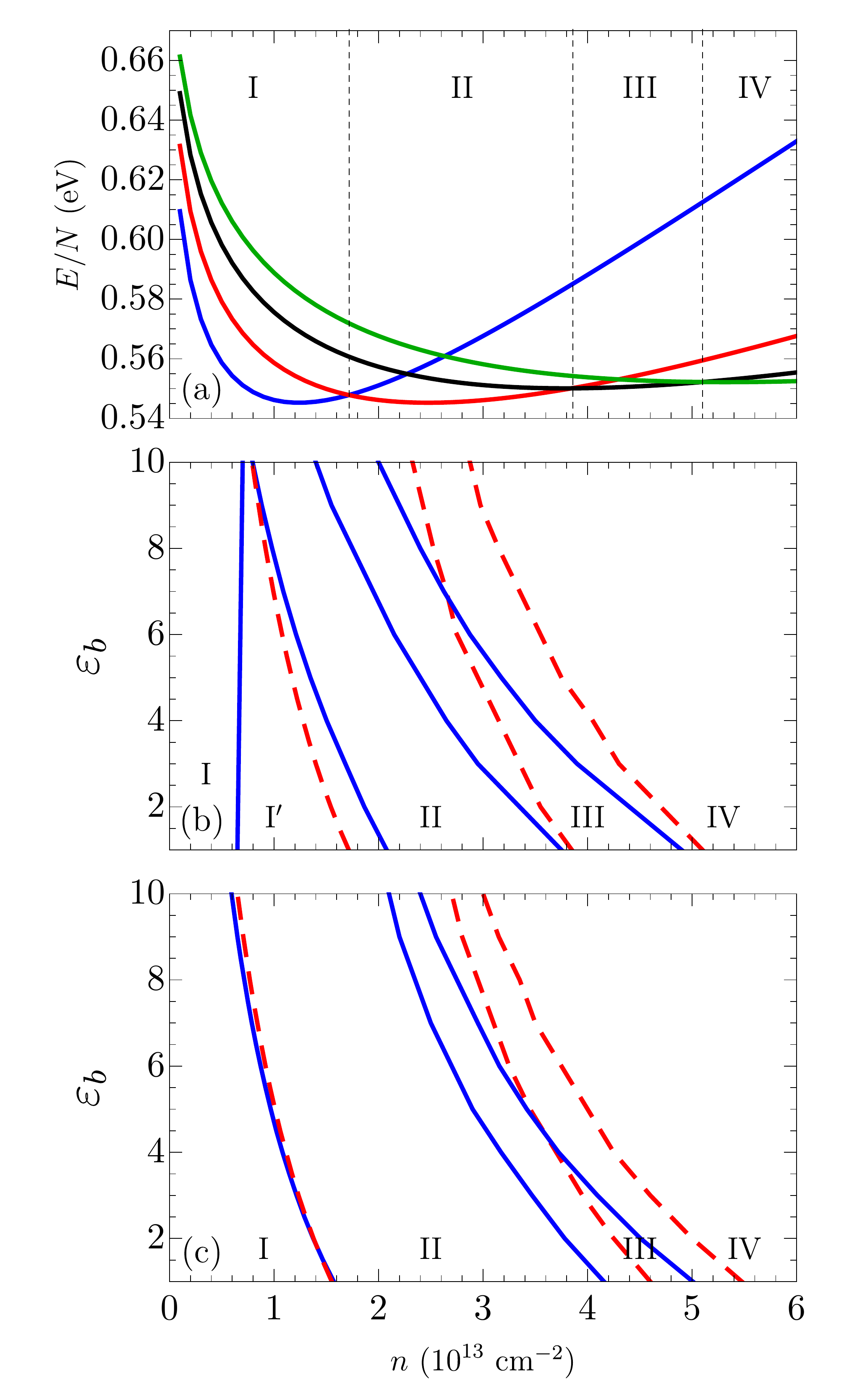}
\caption{(Color online) (a) Energy per particle for MoSe$_2$ for zero magnetic field and $\varepsilon_b=\varepsilon_t=1$ as a function of the total electron density for phase I (blue), phase II (red), phase III (black), and phase IV (green). The phase transitions are indicated by the vertical lines. (b) Phase diagram for zero magnetic field as a function of the total electron density and the dielectric constant of the substrate below the material $\varepsilon_b$ ($\varepsilon_t=1$) for MoS$_2$ (solid, blue) and MoSe$_2$ (dashed, red). (c) The same as (b) but now for WS$_2$ (solid, blue) and WSe$_2$ (dashed, red).}
\label{fig:phasedia}
\end{figure}

When electron-electron interactions are present, we find four different many-body phases as shown in Fig. \ref{fig:phaseplot}. Phase I is characterized by $(\zeta_{\sigma},\zeta_{\tau},\zeta_{\alpha})=(1,1,1)$ or $(\zeta_{\sigma},\zeta_{\tau},\zeta_{\alpha})=(-1,-1,1)$ for Mo-based TMDs and by $(\zeta_{\sigma},\zeta_{\tau},\zeta_{\alpha})=(1,-1,-1)$ or $(\zeta_{\sigma},\zeta_{\tau},\zeta_{\alpha})=(-1,1,-1)$ for W-based TMDs. The system is completely spin polarized and valley polarized, i.e. the many-body state is a truly ferromagnetic state. In the specific case of MoS$_2$ we also find another ferromagnetic phase, phase I$'$, characterized by $(\zeta_{\sigma},\zeta_{\tau},\zeta_{\alpha})=(-1,1,-1)$ or $(\zeta_{\sigma},\zeta_{\tau},\zeta_{\alpha})=(1,-1,-1)$. The difference with phase I is that the electrons now all occupy one of the upper conduction bands as opposed to one of the lower conduction bands. This is possible because of the very small spin-orbit coupling in the conduction band and because the upper conduction bands have a slightly larger effective mass $m_{\sigma,\tau}=\hbar^2\Delta_{\sigma,\tau}/(2a^2t^2)$ reducing their kinetic energy contribution. Phase II is characterized by $(\zeta_{\sigma},\zeta_{\tau},\zeta_{\alpha})=(0,0,1)$ for Mo-based TMDs and by $(\zeta_{\sigma},\zeta_{\tau},\zeta_{\alpha})=(0,0,-1)$ for W-based TMDs, i.e. the low-density phase of the non-interacting case discussed above. There is no global spin and valley polarization but there is spin-valley locking. Phase III is characterized by non-zero values between $-1$ and $1$ for all three variational parameters, which vary as a function of the electron density, such that one of the valleys is completely spin polarized whereas the other valley shows little to no spin polarization. Finally, phase IV is characterized by $(\zeta_{\sigma},\zeta_{\tau},\zeta_{\alpha})=(0,0,\alpha(n))$ for Mo-based TMDs and by $(\zeta_{\sigma},\zeta_{\tau},\zeta_{\alpha})=(0,0,-\alpha(n))$ for W-based TMDs. This is the completely unpolarized phase which was also found in the high density limit without interactions.

The energies of these four phases are shown in Fig. \ref{fig:phasedia}(a) as a function of the electron density. This shows that we find a step by step decrease in the spin/valley order of the many-body state when increasing the density. In the limit of zero density the energy of all these phases converges to $\Delta/2-\lambda_c$, i.e. the lowest single-particle energy. However, at very low densities the system should transit to a Wigner crystal, which is predicted to occur\cite{wigner} at densities of the order of $1\times10^{11}$cm$^{-2}$. Furthermore, we see that the first derivative of the ground state energy shows discontinuities when transiting between phases, meaning that the transitions between these phases are all first order.  In the non-interacting case there are only two phases (phase II and phase IV) with a second order transition between them, but when interactions are included we find an additional phase between them (phase III) and an extra phase at low densities (phase I) with first order transitions between all phases.

The phase diagram as a function of the electron density and the substrate dielectric constant is shown in Figs. \ref{fig:phasedia}(b)-(c). We can see that all the phase transitions occur at lower densities for higher substrate dielectric constants. This is because the substrate weakens the electron-electron interactions. The phase transition between phase I and II is less dependent on the substrate dielectric constant than those between phase II and III and between phase III and IV. The order in which the four different TMDs change phases is different for the phase transition between phase I and II as compared to those between phase II and III and between phase III and IV.
\begin{figure}
\centering
\includegraphics[width=8.5cm]{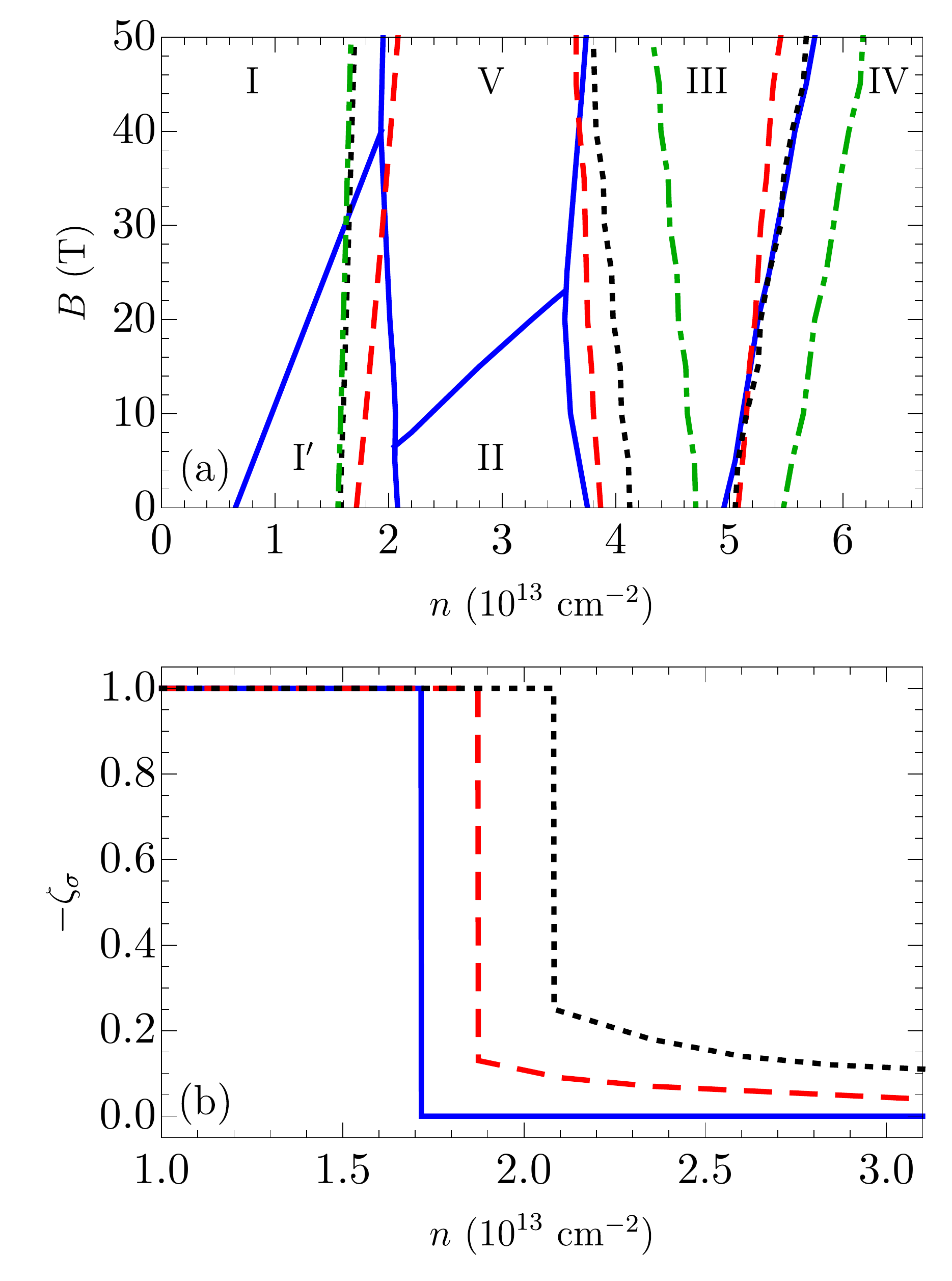}
\caption{(Color online) (a) Phase diagram for $\varepsilon_b=\varepsilon_t=1$ as a function of the total electron density and the perpendicular magnetic field for MoS$_2$ (solid, blue), MoSe$_2$ (dashed, red), WS$_2$ (dotted, black), and WSe$_2$ (dot-dashed, green). (b) Spin polarization for $\varepsilon_b=\varepsilon_t=1$ for MoSe$_2$ as a function of the total electron density for a perpendicular magnetic field of 0 T (solid, blue), 20 T (red, dashed), and 50 T (black, dotted).}
\label{fig:phaseBpol}
\end{figure}

When a perpendicular magnetic field is added to the system, we find the phase diagram shown in Fig. \ref{fig:phaseBpol}(a). The four phases which are present at zero magnetic field persist for non-zero magnetic field. The magnetic field breaks the valley degeneracy and as a result phase I is now only characterized by $(\zeta_{\sigma},\zeta_{\tau},\zeta_{\alpha})=(-1,-1,1)$ for Mo-based TMDs and by $(\zeta_{\sigma},\zeta_{\tau},\zeta_{\alpha})=(1,-1,-1)$ for W-based TMDs. For MoS$_2$ phase I$'$ persists up to magnetic field strengths of 40 T and is characterized by $(\zeta_{\sigma},\zeta_{\tau},\zeta_{\alpha})=(1,-1,-1)$. This means that a complete flip in spin polarization, from $\zeta_{\sigma}=-1$ to $\zeta_{\sigma}=1$, occurs when tuning the system from phase I to phase I$'$. The transition to phase I$'$ occurs at larger densities with increasing magnetic field because the energy difference between the two conduction bands in the lowest energy valley increases with magnetic field. Phase II is now characterized by $(\zeta_{\sigma},\zeta_{\tau},\zeta_{\alpha})=(-\beta(n),-\beta(n),1)$ for Mo-based TMDs and by $(\zeta_{\sigma},\zeta_{\tau},\zeta_{\alpha})=(\beta(n),-\beta(n),-1)$ for W-based TMDs with $\beta(n)$ a function similar to $\alpha(n)$. The exact numerical values of the variational parameters which define phase III and phase IV also change slightly due to the magnetic field but they still represent the same type of phases as shown in Fig. \ref{fig:phaseplot}. We see that the phase transitions between phase I and phase II and between phase III and phase IV shift to higher densities as the magnetic field increases. The phase transition between phase II and phase III, however, shifts to lower densities as the magnetic field increases.

In Fig. \ref{fig:phaseBpol}(b) we show the spin polarization as a function of the electron density. This clearly shows the transition from phase I with complete spin polarization to phase II with partial spin polarization. In the absence of magnetic field the spin polarization in phase II is 0, but this value increases with the magnetic field strength. This also shows that the transition occurs at higher densities for stronger magnetic fields.

\begin{figure}
\centering
\includegraphics[width=8.5cm]{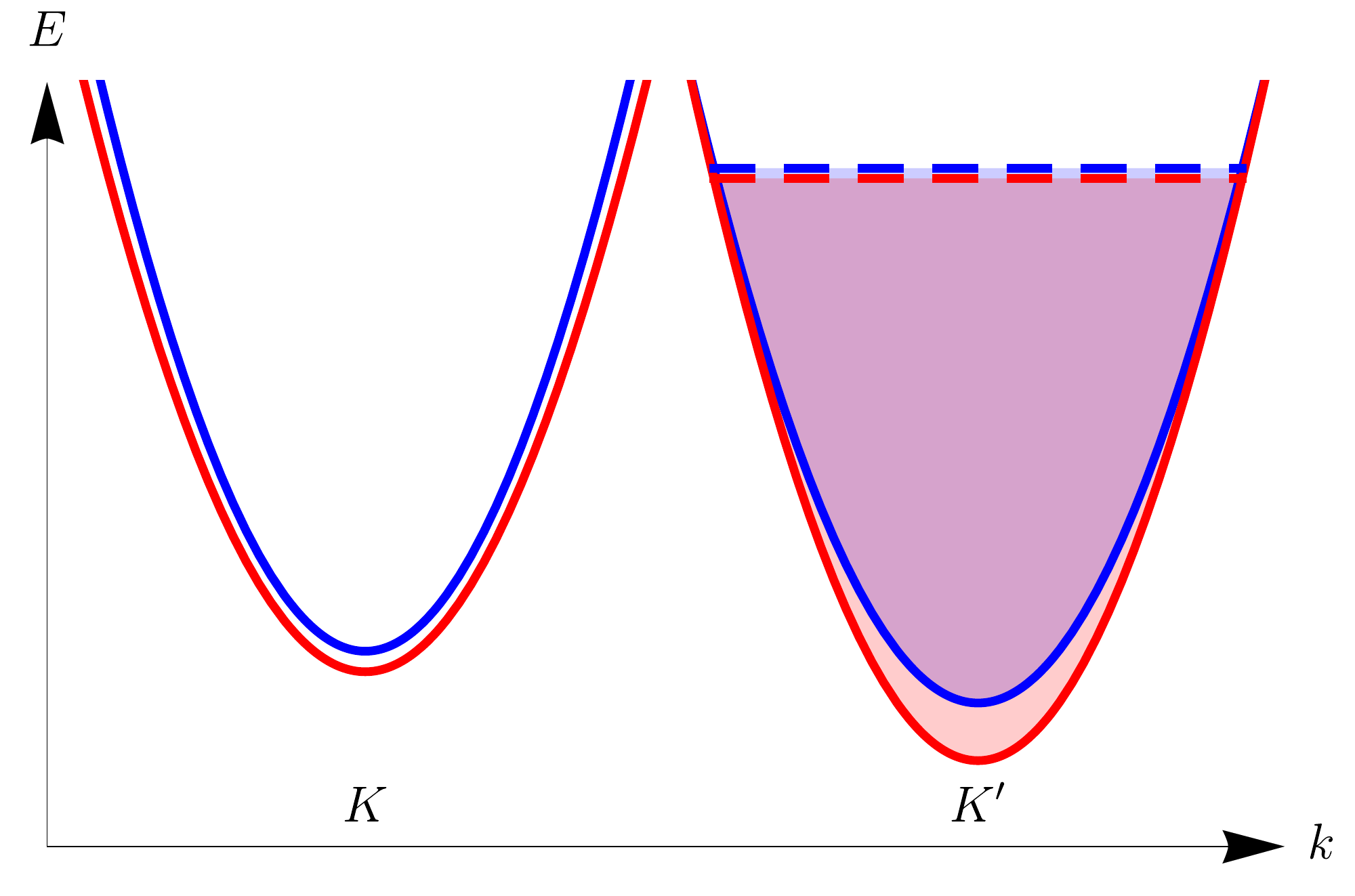}
\caption{(Color online) Phase V in MoS$_2$ when a perpendicular magnetic field is present.}
\label{fig:phaseV}
\end{figure}

Furthermore, for MoS$_2$, we find an additional phase (phase V) at magnetic fields larger than 7 T which completely replaces phase II for magnetic fields larger than 23 T. This phase is characterized by $(\zeta_{\sigma},\zeta_{\tau},\zeta_{\alpha})=(-\beta(n),-1,\beta(n))$ and is shown in Fig. \ref{fig:phaseV}. There is complete valley polarization and very little spin polarization, which is a consequence of the fact that the states in the $K'$ valley shift down in energy with respect to those in the $K$ valley. The reason that we only find this phase for MoS$_2$ is because of the very small spin-orbit coupling in the conduction band. This phase will also occur for the other TMDs but at much stronger magnetic fields.

\subsection{$p$-doped TMDs}

For $p$-doped TMDs we find the same four phases as for $n$-doped TMDs. The transition between phase I and phase II as a function of the substrate dielectric constant is identical to that for $n$-doped TMDs for MoS$_2$ and MoSe$_2$. This is because in both  phase I and phase II only the lowest energy bands are occupied and therefore the energy difference with the higher energy bands due to the spin splitting, which is very different for the conduction and valence bands, has no influence on this phase transition. For WS$_2$ and WSe$_2$, however, we find that this phase transition occurs at lower densities for $p$-doped TMDs as compared to $n$-doped TMDs. The reason is that for these materials the highest valence bands have a smaller effective mass than the lowest conduction bands, whereas for MoS$_2$ and MoSe$_2$ these bands have the same effective mass. Furthermore, we find that phase III and phase IV occur at much higher densities, above $7-15\times10^{13}$ cm$^{-2}$ depending on the TMD and the substrate, as compared to $n$-doped TMDs. This is a consequence of the much stronger spin splitting in the valence bands as compared to the conduction bands. This strong spin splitting also leads to the absence of phase I$'$ for $p$-doped MoS$_2$.

In the presence of a perpendicular magnetic field we find that the transition between phase I and phase II depends more strongly on the magnetic field for $p$-doped TMDs as compared to $n$-doped TMDs. This is a consequence of the coupling of the magnetic field with the magnetic moments of the particles around their atomic site, which only occurs for valence band states. Furthermore, we find that phase V only occurs for unrealistically strong magnetic fields for all TMDs, including MoS$_2$.

\section{Summary and conclusion}
\label{sec:Summary and conclusion}

We used a variational technique to study the different many-body phases of electrons in different monolayer TMDs. We found that there are four phases with first order phase transitions between them. There is a step-wise reduction in spin/valley order with increasing electron density, where the system consecutively exhibits: a complete ferromagnetic phase, complete spin polarization in each of the valleys separately, spin polarization in only one of the valleys, and a paramagnetic phase. We studied the effect of a substrate below the TMD and found that it leads to a reduction in spin/valley order.

Furthermore, we investigated the effect of a perpendicular magnetic field and found that all four phases persist. For the specific case of MoS$_2$ an extra phase appears for magnetic fields larger than 7 T. In this phase there is complete valley polarization but little to no spin polarization. Another effect exclusive to this material is that a complete flip in spin polarization, from $\zeta_{\sigma}=-1$ to $\zeta_{\sigma}=1$, occurs at low densities. Both these effects are the consequence of the very small spin-orbit coupling in the conduction band.

Finally, we also considered $p$-doped TMDs and found that the corresponding phase diagram is less rich than than of $n$-doped TMDs. Phase I$'$ and phase V do not occur and phase III and phase IV only occur at very large densities.

The phase diagrams obtained in the current work could in principle be measured experimentally. Phases with a significant spin polarization, i.e. phases I and III, should be the easiest to observe experimentally by using a magnetometer, although the limited density region in which phase III occurs might hinder observations of this phase. However, to the best of our knowledge such experiments have not yet been reported.

\section{Acknowledgments}

This work was supported by the Research Foundation of Flanders (FWO-Vl) through an aspirant research grant for MVDD and by the FLAG-ERA project TRANS-2D-TMD.

\end{document}